\documentclass[a4paper,10pt]{article}
\usepackage{amssymb,amsmath}
\usepackage{fix-cm}

\begin{document}
\numberwithin{equation}{section}
\begin{flushright}
July 2008 \\
\end{flushright}
\centerline{\bf \fontsize{15}{18}\selectfont Quark-like particles in dual electromagnetic fields}
\bigskip
\centerline{Harry Schiff}
\centerline{\it Professor Emeritus, Department of Physics, University of Alberta,}
\centerline{\it Edmonton,  AB,  Canada, T6G 2J1\footnote{Mailing address: 304-2323 Hamiota St. Victoria BC. CA V8R 2N1\\ \hspace*{5 mm}email: hschiff@shaw.ca}}
\bigskip

\begin{center}
{\bf \underline{Abstract}}\end{center}
\begin{quote}
\hspace*{3 mm} \protect \fontsize{9}{11}\selectfont In a classical field model involving extended dual electromagnetic fields quark-like particles are shown to have fractional charges and a confining energy that provides an asymptotically free spherical surface. A suggestion is made for combining a measure of the confining energy with color in QCD.
\end{quote}
\vskip 2pt

\section{\protect \centering \fontsize{12}{14}\selectfont INTRODUCTION}


\hspace*{5 mm} Recently \cite{sch1} we proposed a quark-like model involving a non-linear extension of Maxwell’s equations. That model did not include a Lagrangian, the purpose of this note is to introduce one. It will be convenient first to summarize the essential features of the model:\\
\hspace*{5 mm} Starting with the observation that for the Li\'enard-Wiechart solutions $F_{\mu\nu}$ and $A_\mu$ for a point charge in arbitrary motion, the mixed field tensor $W_{\mu\nu}\equiv F_{\mu\nu}+\sqrt{2}g^{-1} A_\mu A_\nu$
with charge g satisfies everywhere the null\footnote{Although the term `null' applies strictly to an anti-symmetric tensor we have taken the liberty to use it here \hspace*{5 mm} for the mixed tensor.} conditions
\begin{equation}\label{e1}
W_{\mu\nu}W_{\mu\nu}=0, \qquad {}^*W_{\mu\nu}W_{\mu\nu}=0
\end{equation} 
This led to consideration of the non-gauge invariant equation (e = c = 1),
\begin{subequations}\label{e2}
\begin{align}
 \partial_\mu W_{\mu\nu}&= -4\pi^ej_\nu \label{e2_a}\\
\partial_\mu(F_{\mu\nu}+\sqrt{2}g^{-1} A_\mu A_\nu)&= -4\pi^ej_\nu \label{e2_b}
\end{align}
\end{subequations}
\hspace*{5 mm} Thus the total conserved current consists of the localized external current ${}^e j_\nu$
plus the field contribution $\sqrt{2}(4\pi g)^{-1}\partial_\mu(A_\mu A_\nu)$. As shown below it is the latter in combination with external charges that allows for the realization of fractional charges.\\
\hspace*{5 mm} Simple static and radially symmetric solutions of (\ref{e2}) to which the null conditions (\ref{e1}) are applied are considered below.

\section{\protect \centering \fontsize{12}{14}\selectfont  SOLUTIONS OF (\ref{e2_b})}

\hspace*{5 mm} For radially symmetric static solutions of (\ref{e2_b}), where the null condition
${}^*W_{\mu\nu}W_{\mu\nu}=\textbf{E}\cdot \textbf{B}=0$ is identically satisfied, involve the equations,
\begin{eqnarray}
\sqrt{2} g^{-1}\nabla\cdot(A\textbf{A})= -4\pi^eJ(r) \label{e3}\\
\nabla\cdot \textbf{E} - \sqrt{2} g^{-1}\nabla\cdot(\phi\textbf{A}) = 4\pi^e\rho(r) \label{e4}
\end{eqnarray} 
The asymptotic solution of (\ref{e3}), from the homogeneous equation
\begin{equation}
\nabla\cdot(A\textbf{A})=0
\end{equation} 
is
\begin{equation}
 A= a/r \qquad \textrm{(a arbitrary)}
\end{equation}
Defining
\begin{equation}
 \gamma\equiv \sqrt{2}g^{-1}a,
\end{equation}
\begin{equation}\label{e5a}
 A=\gamma g/\sqrt{2}r
\end{equation}

With (\ref{e5a}) the homogeneous part of (\ref{e4}) can be written,
\begin{equation}\label{e4a}
 \frac{d^2(r\phi)}{rdr^2}+\frac{\gamma d(r\phi)}{r^2 dr}=0
\end{equation}
For any value of $\gamma$ except $\gamma=1$ treated separately below there are two indicial solutions of (\ref{e4a}),
\begin{equation}\label{e7}
 \phi_1 = b/r
\end{equation}
\begin{equation}\label{e8}
 \phi_2 = cr^{-\gamma}
\end{equation}

consisting of a Coulomb potential and for $\gamma<0$ a possible absolute confining potential, 
where c may be positive or negative. In the following the particular solution to (\ref{e4}) for which (\ref{e7}) is the asymptotic solution will be called $\phi_p$ . For any value of $\gamma$ the total charge of the source in (\ref{e4}) using (\ref{e7}), (\ref{e5a}) and applying the divergence theorem is
\begin{equation}\label{e29}
 \int {}^e\rho d^3x=(1-\gamma)b
\end{equation} 
 \hspace*{5 mm} Using the divergence theorem in (\ref{e3}) the total 3-current of the source is
\begin{equation}\label{e210}
 \int {}^eJ(r)d^3x=-1/4\pi \int \sqrt{2}g^{-1}A^2r^2d\Omega =-\gamma^2 g/\sqrt{2}
\end{equation} 
\hspace*{5 mm} For $\gamma=1$ the two indicial solutions of (\ref{e4a}) merge to a single Coulomb solution so a second solution is needed. This is easily seen to be given by  $r\phi\sim \log r$. These solutions however will not be considered here any further.

\section{\protect \centering \fontsize{12}{14}\selectfont  NULL CONDITION AND FRACTIONAL CHARGES }
\hspace*{5 mm} Explicitly the null condition $W_{\mu\nu}=0$ is
\begin{equation}\label{e31}
 F_{\mu\nu}F_{\mu\nu}+2g^{-2}(A_\nu A_\nu)^2=0
\end{equation} 
The radially symmetric asymptotic solutions of (\ref{e31}) gives
\begin{equation}
 E^2+g^{-2}(A^2-\phi^2_1)^2=0
\end{equation} 
Use of (\ref{e5a}) and (\ref{e7}) yields a quadratic equation for the Coulomb charge b in (\ref{e7}),
\begin{equation}\label{e32}
 b^2\pm gb - \gamma^2 g^2 /2 =0
\end{equation} 

Solutions b- with the minus sign in (\ref{e32}) have opposite signs to those with the plus sign b corresponding to $g \rightarrow -g$. Thus the quadratic
\begin{equation}
 b^2 -gb-\gamma^2 g^2 /2=0
\end{equation} 
has the two solution for $b_-$,
\begin{equation}\label{e34}
 b_-(\pm)=g\left[ 1\pm \sqrt{1+2\gamma^2} \right] /2
\end{equation} 
In (\ref{e34}) choosing $\gamma^2=4$ and $g=\pm 1/3$, on gets
\begin{equation}
 b_-(\pm)=\pm (2/3, -1/3), \quad b_+=-b_-
\end{equation} 
\hspace*{5 mm} With $b=\pm1/3, \pm2/3$ and $\gamma=-2$ one finds that the external charge (\ref{e29}) is $\pm1,\pm2$ respectively. (Other possible fractional values for b can be obtained, for example for $\gamma^2=4$ and $g=\pm2/3, b_-=\pm(4/3, -2/3), b_+=-b_-$).\\
Choosing (\ref{e8}) to be a confining potential the value $\gamma=-2$ is used below.

\section{\protect \centering \fontsize{12}{14}\selectfont LAGRANGIAN }
 \hspace*{5 mm} A Lagrangian for (\ref{e2_b}) cannot be obtained using only the field components in (\ref{e2_b}), a supplemental field is required. To this end an additional electromagnetic field $G_{\mu\nu}$ is chosen with its associated potential $V_\mu$,
\begin{equation}
 G_{\mu\nu}=\partial_\mu V_\nu - \partial_\nu V_\mu
\end{equation} 
The following Lagrangian includes $F_{\mu\nu}$, $A_\nu$ and $G_{\mu\nu}$, $V_\nu$ with their respective currents ${}^ej_\nu$ and ${}^*j_\nu$.
\begin{equation}\label{e42}
 \mathcal L = -1/8\pi \left[ 1/2 G_{\mu\nu}F_{\mu\nu} + \sqrt{2}g^{-1}V_{\nu}\partial_\mu (A_\mu A_\nu) + 4\pi V_\nu {}^ej_\nu + 4\pi A_\nu {}^*j_\nu \right] 
\end{equation} 
\hspace*{5 mm} Variation with respect to $V_\mu$ gives (\ref{e2_b}), while variation with respect to $A_\mu$ yields a linear equation for $V_\mu$. 
\begin{equation}\label{e43}
 \partial_\mu G_{\mu\nu} - \sqrt{2}g^{-1}A_\mu \left( \partial_\mu V_\nu + \partial_\nu V_\mu \right) =-4\pi^*j_\nu 
\end{equation} 
The total current
\begin{equation}
 j_\nu=-\sqrt{2}(4\pi g)^{-1} A_\mu \left( \partial_\mu V_\nu + \partial_\nu V_\mu \right) + {}^*j_\nu
\end{equation} 
is conserved and for radially symmetric solutions considered here the 3-current $J=0$.
\\
 \hspace*{5 mm} From (\ref{e43}) the equation for $V_4=i\Psi$ becomes
\begin{equation}\label{e45}
 \frac{d^2\Psi}{dr^2} + \frac{2d\Psi}{r dr} - \sqrt{2}g^{-1}A\frac{d\Psi}{dr} = -4\pi^*\rho(r)
\end{equation} 
Both particular and homogeneous solutions of (\ref{e45}) will be considered. Consider first the homogeneous $\Psi_h$.
\begin{equation}\label{e46}
 \frac{d^2\Psi_h}{dr^2} + \frac{2d\Psi_h}{r dr} - \sqrt{2}g^{-1}A\frac{d\Psi_h}{dr} = 0
\end{equation} 
Using the asymptotic value of A in (\ref{e5a}) with $\gamma=-2$, (\ref{e46}) becomes
\begin{equation}
 \frac{d^2\Psi_h}{dr^2} + \frac{4d\Psi_h}{r dr}=0
\end{equation} 
Thus asymptotically,
\begin{equation}
 \Psi_h \sim 1/r^3
\end{equation} 
Using a simple expression for $\Psi_h$ in (\ref{e46}) such as
\begin{equation}\label{e49}
 \Psi_h = \lambda s^2/(s+r)^3
\end{equation} 
with arbitrary charge $\lambda$ and a scale factor $s>0$ one obtains for the 3-potential A,
\begin{equation}\label{e410}
 \sqrt{2}g^{-1}A= \left[ 2/r - 4/(s+r) \right] 
\end{equation} 
\hspace*{5 mm} With (\ref{e410}) in (\ref{e4}) one obtains the homogeneous solution for the confining potential corresponding to asymptotic (\ref{e8}),
\begin{equation}\label{e411}
 \phi_k = k (s+r)^4/s^3r^2
\end{equation} 
where k is arbitrary with dimension of charge ($k/s^3$ replaces c in (\ref{e8})). Substituting (\ref{e49}) in (\ref{e3}) one finds for the external 3-current density
\begin{equation}\label{e412}
 {}^eJ(r)=-2 \sqrt{2}g \left[ \delta(r) -s(s-r)/\pi^2(s+r)^3 \right] 
\end{equation} 
Spatial integration of (\ref{e412}) is in accordance with (\ref{e210}).\\
\hspace*{5 mm} For the 3-potential V in (\ref{e43}) we choose the external 3-current ${}^*J=0$ so the equation for V becomes
\begin{equation}\label{e413}
 2A\frac{dV}{dr}=\phi_p \frac{d\Psi_h}{dr}
\end{equation} 
From the behavior of A in (\ref{e410}), $d\Psi_h/dr$ from (\ref{e49}) and $\phi_p$ at the origin and asymptotically, (taking $\phi_p(0) = \textrm{const} $), $V\sim r^2$ and $1/r^3$ respectively. A simple choice for V is then
\begin{equation}
 V= \tau s^2 r^2 / (s+r)^5
\end{equation} 
with arbitrary charge $\tau$. The particular solution $\phi_p$ of (\ref{e4}) follows from (\ref{e413}),
\begin{equation}
 \phi_p = -2 \sqrt{2}g\tau(3\lambda)^{-1} \left[ 2/(s+r) - 9r/(s+r)^2 + 10r^2/(s+r)^3 \right] 
\end{equation} 
Since $\phi_p \rightarrow b/r$ asymptotically this requires that
\begin{equation}
 -2\sqrt{2}g\tau/\lambda =b
\end{equation} 
Thus
\begin{equation}\label{e415a}
 \phi_p=b/3\left[ 2/(s+r) - 9r/(s+r)^2 + 10r^2/(s+r)^3 \right] 
\end{equation} 
\hspace*{5 mm} For the particular solution $\Psi_p$ of (\ref{e45}) we choose the same function as (\ref{e415a}) allowing the associated charge ${}^*b$ to be different from the charge b in $\phi_p$,
\begin{equation}\label{e417}
 \Psi_p = {}^*b/3\left[ 2/(s+r) - 9r/(s+r)^2 + 10r^2/(s+r)^3 \right]
\end{equation} 
Applied to (\ref{e45}) one obtains the total charge density 
\begin{equation}
 {}^*\rho_t(r)=\sqrt{2}(4\pi g)^{-1}A d\Psi_p/dr +{}^*\rho(r) = \frac{{}^*b(14r^2s-35rs^2+11s^3)}{(6\pi)r(s+r)^5}
\end{equation} 
The complete solution $\Psi=\Psi_p + \Psi_h$ to (\ref{e45}) using (\ref{e49}), (\ref{e417}) and replacing $\lambda$ by ${}^*\lambda$,
\begin{equation}
 \Psi = {}^*b/3\left[ 2/(s+r) - 9r/(s+r)^2 + 10r^2/(s+r)^3 \right] + {}^*\lambda s^2/(s+r)^3
\end{equation} 

\section{\protect \centering \fontsize{12}{14}\selectfont MASS }
\hspace*{5 mm} From the Lagrangian (\ref{e42}) the canonical stress-energy tensor is
\begin{equation}\label{e51}
 T_{\alpha\beta}=-\partial \mathcal L \frac{\partial_\beta A_\lambda}{\partial(\partial_\alpha A_\lambda)} - \partial \mathcal L \frac{\partial_\beta V_\lambda}{\partial(\partial_\alpha V_\lambda)}
+ \mathcal L \delta_{\alpha\beta}
\end{equation}
To represent the energy of a particle we choose, for simplicity, the stress-energy tensor of the electromagnetic fields of (\ref{e51}) involving $F_{\alpha\mu}$ and $G_{\alpha\mu}$
\begin{equation}
 \Theta_{\alpha\beta} = 1/8\pi \left[ G_{\alpha\mu}F_{\beta\mu} + F_{\alpha\mu}G_{\beta\mu} -1/2(G_{\mu\nu}F_{\mu\nu})\delta_{\alpha\beta} \right] 
\end{equation} 
with energy density
\begin{equation}\label{e53}
 -\Theta_{44} = 1/8\pi \left[-2G_{4\mu}F_{4\mu} + 1/2G_{\mu\nu}F_{\mu\nu} \right] 
\end{equation}
\hspace*{5 mm} For two different charges ${}^*b$ and $b$ as well as different ${}^*\lambda$ and $\lambda$, (\ref{e53})  can be written in two ways, with ${}^*b$ in $\Psi_p$, $b$ in $\phi_p$ and vice versa. Excluding the confining potential $\phi_k$ (\ref{e411}) for now the two different energy densities are,

\begin{subequations}\label{e54}
\begin{align}
-{}^*\Theta_{44} = 1/8\pi \frac{d}{dr}\left[ \Psi_p({}^*b) + \Psi_h({}^*\lambda) \right]  \frac{d\phi(b)}{dr}\\
-\Theta_{44} = 1/8\pi \frac{d}{dr}\left[ \Psi_p(b) + \Psi_h(\lambda) \right]  \frac{d\phi_p({}^*b)}{dr}
\end{align}
\end{subequations}
The products $[d\Psi_p({}^*b)/dr][d\phi_p(b)/dr]$ and $[d\Psi_p(b)/dr][d\phi_p({}^*b)/dr]$ in (\ref{e54}) suggest interactions between the two charges, however these are not expressions for usual interactions. By replacing ${}^*b$ with $b$ in (\ref{e54}) and ${}^*\lambda$ by $\lambda$ the two energy densities result in the expression for the energy density of a single charge, noting that now $\Psi_p = \phi_p$. Changing $-\Theta_{44}$ to $-\Theta^{'}_{44}$,
\begin{equation}
 -\Theta^{'}_{44}=\frac{1}{8\pi} \left[ \frac{d\phi_p(b)}{dr}\frac{d\phi_p(b)}{dr} + \frac{d\Psi_h(\lambda)}{dr}\frac{d\phi_p(b)}{dr} \right] 
\end{equation} 
with integrated energy
\begin{equation}\label{e56}
 \xi'=\frac{1}{2}\int \left[ \frac{d\phi_p(b)}{dr} \frac{d\phi_p(b)}{dr} + \frac{d\Psi_h(\lambda)}{dr} \frac{d\phi_p(b)}{dr} \right] r^2 dr 
\end{equation}
In (\ref{e56}) the first term is a Coulomb energy while the second is a van der Waals type
energy in which $d\Psi_h(\lambda)/dr$ goes as $1/r^4$ asymptotically (see (\ref{e49})) and $d\phi_p(b)/dr$ goes as $1/r^2$.  We interpret this as a binding energy for equal $\Psi_p$ and $\phi_p$ in the expression for the Coulomb energy. Using the functions (\ref{e415a}) and (\ref{e49}), integration of (\ref{e56}) yields,
\begin{equation}\label{e57}
 \xi'= 0.0391b^2 + 0.386\lambda b/s
\end{equation} 
\hspace*{5 mm} To include the confining energy we adopt a `bag' model assuming zero interactions between particles and take the confining energy for $a$ charge $b$ to be given by $b\phi_k$. With (\ref{e57}) the energy
\begin{equation}
 \xi = \xi' + \xi_k
\end{equation} 
where
\begin{equation}
 \xi_k = b\phi_k = bk (s+r)^4/s^3r^2
\end{equation} 
$\xi_k>0$ for $bk>0$ and represents a well with sides increasing as $1/r^2$ as $r\rightarrow 0$ and as $r^2$ asymptotically. The minimum occurs at $r = s$ where $\xi_{kmin} =  16bk/s$. On the spherical surface $r = s$ particles move force-free, an area of `asymptotic freedom'. The `size' of the bag would be determined by an r value corresponding to quark separation energy.
The minimum energy or mass is thus
\begin{equation}\label{e510}
 m = \xi_{min} = (0.0391b^2 +0.386\lambda b + 16bk)/s
\end{equation} 
with parameters $\lambda$ and $k$ arbitrary for a given charge $b$. A binding van der Waals
energy would require $\lambda b < 0 $ in contrast to $bk>0$ so that $\lambda$ and $k$ would have opposite signs.

\section{\protect \centering \fontsize{12}{14}\selectfont COLOR}
\hspace*{5 mm} The expression for the mass in (\ref{e510}) has three parameters, $k$, $\lambda$ and $s$ as well as two numerical quantities that are solution dependant; the equations allows for other possible solutions. This form of the energy raises the possibility where two particles may have equal charges and equal masses, $m$ and ${}^*m$, with different pairs of parameter values $\lambda, k$ and ${}^*\lambda, {}^*k$ such as these non-identical twins
\begin{subequations}\label{e61}
\begin{align}
m = (0.0391b^2 + 0.386\lambda b + 16bk)/s  \label{e61a}\\
{}^*m = (0.0391b^2 + 0.386{}^*\lambda b + 16b^*k)/s \label{e61b}
\end{align}
\end{subequations}
\hspace*{5 mm} In another context one is also aware of twin masses such as the two up-quarks in a proton and the two down-quarks in a neutron which are distinguished, for reason of symmetry, by a color designation that extends also to the gluons that provide the strong force on quarks. \\
\hspace*{5 mm} Having trod this far into the QCD arena, another step is tempting within the limitations of this static model, to combine it with that of color in QCD. To this end we choose the charge products bk of the confining energy.  In particular, the following is proposed: For the traditional up and down-quarks usually shown as u, d and anti-quarks here as $\underline{u}$, $\underline{d}$, the corresponding bk products are written as $uk_u$ $dk_d$ and $\underline{uk}_u$, $\underline{dk}_d$ where $\underline{k}$ is the confining charge for an anti-quark. The $k$ charges of twins are shown as $k$ and ${}^*k$. The three colors, red, green and blue are identified by 3 positive indefinite numbers R, G and B and their negative anti-colors by $\underline{R}, \underline{G}$, and $\underline{B}$. The \textit{quantity} of color of a particular quark is expressed as $Rbk, Gbk,$ or $Bbk$ (all positive for $bk>0$). These are referred to as color-contents. According to QCD the color!
 s of quarks change continually so that $R, G$ and $B$ will move continually between the bk values; as a requirement of color neutrality they may change in magnitude as they reach different quarks and their bk values.\\
\hspace*{5 mm} Consider first a charged meson with, say, a red u-quark and an anti-red anti-d-quark. For an expected total color-content of zero the individual color-contents combine as
\begin{equation}\label{e62}
 Ruk_u + \underline{Rdk}_d = 0
\end{equation} 
with $\underline{k}_d > 0$ (all $bk>0$). For unequal $bk$ values in (\ref{e62}) the $R$ and $\underline{R}$ magnitudes would be unequal.\\
For the two u-quarks and one d-quark in a proton their three positive color-contents  combine as
\begin{equation}\label{e63}
 Ruk_u + Gu^*k_u + Bdk_d = W_p
\end{equation} 
with $W_p$ standing for white. The three RGB values have different magnitudes in general and move continually between the different bk values while maintaining color neutrality. For a color neutrality of white a `palette' consisting of 3 equal color-contents in (\ref{e63}) results in the equalities, 
\begin{equation}
 Ruk_u = Gu^*k_u = Bdk_d
\end{equation} 
Of note are the equal color-contents of the twin u-quarks where their color difference is reflected in the unequal $k$ charges, $k_u$ and ${}^*k_u$. The only apparent physical distinction here is in their different confining energies, $2/3\phi_k$. For a neutron with one u-quark and two d-quarks, their color-contents combine for color neutrality as
\begin{eqnarray}
 Ruk_u + Gdk_d +Bd^*k_d = W_n&& \\
 \textrm{with} \qquad Ruk_u = Gdk_d =Bd^*k_d&&
\end{eqnarray} 
Similarly to the proton the equal color-contents of the twin d-quarks show their color difference reflected in the unequal $k$ charges, $k_d$ and ${}^*k_d$.\\
\hspace*{5 mm} The u-quark in (6.5) can be attached either to $k_u$ or ${}^*k_u$ which one it may be is probably random. Similar comments apply to the d-quark in the proton with respect to $k_d$ and ${}^*k_d$. In general, $W_n$ may be different from $W_p$. The $RGB$ magnitudes in the neutron may not be the same as those in the proton, so for example $Ruk_u$(proton) may not be equal to $Ruk_u$ (neutron) or $Ru^*k_u$ (neutron). Except for their relative magnitudes the actual values of $R, G,$ and $B$ are irrelevant here.\\
\hspace*{5 mm} According to the above scheme for combining this model with color in QCD, the u-quarks, d-quarks and anti-quarks that are involved in the structures of the proton, neutron and meson families adjoin 8 confining k charges: ( $k_u, {}^*k_u, k_d, {}^*k_d$) and anti-charges ($\underline{k}_u, {}^*\underline{k}_u, \underline{k}_d, {}^*\underline{k}_d$), matching the number of 8 different gluons.
\section{\protect \centering \fontsize{12}{14}\selectfont CONCLUSIONS}
This model describes the mass-energy of quarks interacting with 8 confining charges that are identified with 8 gluons. The interactions appear in the confining energy and in the contributions to the quark mass by the interaction term 16bk/s (\ref{e510}).  The mass-energy of gluons is not involved in the model so, for example, the gluon mass is not present.\\
\hspace*{5 mm} The choice of the particular Lagrangian (\ref{e42}) is not conventional, missing as it does the usual kinetic terms $F_{\mu\nu}F_{\mu\nu}$ and $G_{\mu\nu}G_{\mu\nu}$. This is overcome to the degree that the choice of equal particular solutions of the two field equations merges $F_{\mu\nu}$ and $G_{\mu\nu}$ and identifies the Coulomb energy term of a single charge shown in (\ref{e56}). In addition to the occurrence of fractional charges $b = \pm(2/3, -1/3)$ the homogeneous  solution for the confining potential $\phi_k$ leads to an asymptotically free spherical surface on which the charges can move freely without interaction. A second homogeneous solution contributes an additional energy of a characteristic van der Waals type.\\
\hspace*{5 mm} These results based on the dual modified electromagnetic fields present a different and possibly fruitful perspective on QCD in spite of being incomplete, with spin and kinetic energy absent.  An open question is the unusual consequence, referred to above, where the identity of a single charge arises from the contributions of two different equations. Possible implications for an inner quark structure have not escaped our notice.

\section*{\protect \centering \fontsize{12}{14}\selectfont ACKNOWLEDGMENT}
I want to thank J. David Jackson for helpful comments and suggestions also Werner Israel for useful discussion

\pagebreak

\end{document}